\documentstyle[aps,twocolumn,epsfig]{revtex}
\begin{document}
\draft

\twocolumn[\hsize\textwidth\columnwidth\hsize\csname@twocolumnfalse\endcsname

\title{Coherence and squeezing in superpositions of spin coherent states}
\author{Xiaoguang Wang}
\address{Institute of Physics and Astronomy, Aarhus University, DK-8000, Aarhus C, Denmark, and}
\address{Quantum information processing group, 
Institute for Scientific Interchange (ISI) Foundation, Viale Settimio Severo 65, 
I-10133 Torino, Italy}

\date{\today}
\maketitle

\begin{abstract}
We consider the superpositions of spin coherent states and study 
the coherence properties and spin squeezing in these states. 
The spin squeezing is examined using a new version of spectroscopic 
squeezing criteria. The results show that the antibuching effect
can be enhanced and spin squeezing can be generated in the
superpositions of two spin coherent states. 
\end{abstract}

\pacs{Keywords: Coherence, spin squeezing and spin coherent state.}
]

\section{Introduction}

There has been much interest in the study of Schr\"{o}dinger cat states in
quantized electromagnetic field\cite{EM} and quantized motion of the center
of mass of  a trapped ion\cite{Ion}. Two types of cat states, superpositions
of the bosonic coherent state $|\alpha \rangle $ and $|-\alpha \rangle $\cite
{EM} or superpositions of $|\alpha \rangle $ and $|\alpha ^{*}\rangle $\cite
{RI}, are studied in the literature. These states exhibit nonclassical 
properties such as oscillations of photon number distribution, antibunching
effects and quadrature squeezing.

Now we replace the bosonic coherent state $|\alpha \rangle $ by the spin
coherent state (SCS) \cite{SCS} $|\eta \rangle $ in the cat states and
obtain a superposition of two SCSs. Agarwal {\it et al}.\cite
{Agar97}, Gerry {\it et al}.\cite{Gerry97} and Recamier {\it et al}.\cite
{Recamier} have introduced and studied the generation scheme and some properties of
the superposition state. Here we concentrate on their coherence properties and spin
squeezing\cite{Kitagawa}. There are many ways to characterize spin squeezing%
\cite{Kitagawa,Wineland94,Bg}. Here we use the squeezing criteria given by
Wineland {\it et al}.\cite{Wineland94} and S\o rensen {\it et al.}\cite
{Soerensen3}. 
The squeezing parameter is defined as
\begin{equation}
\xi _{\vec{n}_1}^2={\frac{N(\Delta J_{\vec{n}_1})^2}{\langle J_{\vec{n}%
_2}\rangle ^2+\langle J_{\vec{n}_3}\rangle ^2}},  \label{eq:xi}
\end{equation}
where $J_{\vec{n}}=\vec{n}\cdot \vec{J},$ $\vec{n}_i(i=1,2,3)$ are
orthogonal unit vectors and $\vec{J}$ is the angular momentum operator. The
states with $\xi _{\vec{n}}^2<1$ are spin squeezed in the direction $\vec{n}$. 
Under this criteria one remarkable feature is that the squeezing parameters $\xi
_x^2=\xi _y^2=\xi _z^2=1$ \cite{Wangspin} for spin coherent states \cite{SCS}%
. Another reason to use this criteria is that the squeezing parameter defined
in Eq.(\ref{eq:xi}) is closely related to the entanglement of many qubits \cite{Soerensen3}, i.e., 
the condition $\xi _{\vec{n}}^2<1$ also indicates entanglement
in multi-qubit systems. 
It is well-known that quadrature squeezing in bosonic systems can be generated in nonlinear Kerr medium. 
Similar nonlinear Hamiltonian was proposed by Kitagawa and Ueda \cite{Kitagawa} to produce 
spin squeezing. It is also known that the quadrature squeezing exist in the cat states, then it is reasonable
to examine the spin squeezing in the superpositions of spin coherent states. 
So the present works on coherence properties and spin squeezing in spin systems are 
complementary to  the previous works on coherent properties and quadrature squeezing
in bosonic systems.

The paper is organized as follows: In sec. II, we introduce the superposition
of two SCSs and give its corresponding ladder operator formalism which is useful in the
calculation of the expectation value $\langle J_-^2\rangle $. We examine the
coherence properties in Sec. III and spin squeezing in Sec. IV. Finally a
conclusion is given in Sec. V.

\section{The SSCS and its ladder operator formalism}

We work in the $(2j+1)$-dimensional angular momentum Hilbert space $%
\{|j,m\rangle ;m=-j,...,+j\}$. It is convenient to define the `number'
operator ${\cal N}=J_z+j$ and `number states' as 
\begin{eqnarray}
|n\rangle  &\equiv &|j,-j+n\rangle ,  \nonumber \\
{\cal N}|n\rangle  &=&n|n\rangle .
\end{eqnarray}
The SCS is defined in this Hilbert space and given by\cite{SCS}, 
\begin{equation}
|\eta \rangle =(1+|\eta |^2)^{-j}\sum_{n=0}^{2j}{{%
{2j \choose n}%
}}^{1/2}\eta ^n|n\rangle,
\end{equation}
where the parameter $\eta $ is complex. The inner product $\langle \eta
|-\eta \rangle ={\xi }^{2j},\,$where $\xi =\frac{1-|\eta |^2}{1+|\eta |^2}$
is a real number. Formally the SCS is exactly of the form of the binomial
state. \cite{BS}

The quantum state, which is of interest here, is the superposition of two SCS $%
|\pm \eta \rangle $, the SSCS 
\begin{eqnarray}
|\eta ,\theta \rangle  &=&\frac 1{\sqrt{2+2\cos \theta \xi ^{2j}}}\left(
|\eta \rangle +e^{i\theta }|-\eta \rangle \right)   \nonumber \\
&=&\frac{(1+|\eta |^2)^{-j}}{\sqrt{2+2\cos \theta \xi ^{2j}}} \nonumber\\
&&\times\sum_{n=0}^{2j}{%
{%
{2j \choose n}%
}}^{1/2}\eta ^n[1+e^{i\theta }(-1)^n]|n\rangle ,  \label{eq:state}
\end{eqnarray}
where $\theta $ is a relative phase.

It is easy to check that a ladder operator formalism of the SCS is

\begin{equation}
J_{-}|\eta \rangle =\eta (2j-{\cal N})|\eta \rangle ,  \label{eq:laddercs1}
\end{equation}
where $J_{\pm }=J_x\pm iJ_y$. From Eqs.(\ref{eq:state}) and (\ref
{eq:laddercs1}), we find that the SSCS satisfies

\begin{equation}
J_{-}|\eta ,\theta \rangle =\eta \sqrt{\frac{1-\cos \theta \xi ^{2j}}{1+\cos
\theta \xi ^{2j}}}(2j-{\cal N})|\eta ,\theta +\pi \rangle
\label{eq:laddercs2}
\end{equation}
and

\begin{equation}
J_{-}^2|\eta ,\theta \rangle =\eta ^2(2j-{\cal N})(2j-{\cal N}-1)|\eta
,\theta \rangle .  \label{eq:laddercs3}
\end{equation}
Eq.(\ref{eq:laddercs3}) gives the ladder operator formalism of the SSCS $%
|\eta ,\theta \rangle $. 

A specific SSCS $|\eta,\pi/2\rangle$ can be generated in the nonlinear Hamiltonian system 
$H=\chi {\cal N}^2$\cite{Kitagawa} since
\begin{equation}
|\eta,\pi/2\rangle=\exp(-i\pi{\cal N}^2/2)|\eta\rangle
\end{equation}
up to a trivial global phase. Next we begin our discussion on the coherence
properties and spin squeezing in the SSCS.

\section{Coherence properties}

The coherence properties of a spin state can be characterized by the
normalized second-order correlation function\cite{G2}

\begin{eqnarray}
g^{(2)} &=&\frac{\langle J_{+}^2J_{-}^2\rangle }{\langle J_{+}J_{-}\rangle ^2%
}  \nonumber \\
&=&\frac{\langle {\cal N}({\cal N}-1)(2j-{\cal N}+1)(2j-{\cal N}+2)\rangle }{%
\langle {\cal N}(2j-{\cal N}+1)\rangle ^2}.  \label{eq:gggg}
\end{eqnarray}
Then we need to know the expectation values $\langle {\cal N}^k\rangle
(k=1...4),$ which can be conveniently obtained by the generation function
method. The generation function of the SSCS is directly calculated as

\begin{eqnarray}
G(\lambda ) &=&\langle \eta ,\theta |\lambda ^{{\cal N}}|\eta ,\theta
\rangle   \nonumber \\
&=&\frac{\left( 1+\lambda |\eta |^2\right) ^{2j}+\cos \theta \left(
1-\lambda |\eta |^2\right) ^{2j}}{\left( 1+|\eta |^2\right) ^{2j}+\cos
\theta \left( 1-|\eta |^2\right) ^{2j}}
\end{eqnarray}
The factorial moments follow from the generation function 
\[
F(k)={\frac{d^kG(\lambda )}{d^k\lambda }}|_{\lambda =1}=\frac{(2j)!}{(2j-k)!}%
|\eta |^{2k}
\]
\begin{equation}
\times \frac{(1+|\eta |^2)^{2j-k}+(-1)^k\cos \theta \left( 1-|\eta
|^2\right) ^{2j-k}}{\left( 1+|\eta |^2\right) ^{2j}+\cos \theta \left(
1-|\eta |^2\right) ^{2j}}  \label{eq:ffff}
\end{equation}

Relations between the expectation values $\langle {\cal N}^k\rangle $ and
the factorial moments $F(k) (k=1...4)$ are given by the equation

\begin{equation}
\left( 
\begin{array}{l}
\langle {\cal N}\rangle \\ 
\langle {\cal N}^2\rangle \\ 
\langle {\cal N}^3\rangle \\ 
\langle {\cal N}^4\rangle
\end{array}
\right) =\left( 
\begin{array}{llll}
1 & 0 & 0 & 0 \\ 
1 & 1 & 0 & 0 \\ 
1 & 3 & 1 & 0 \\ 
1 & 7 & 6 & 1
\end{array}
\right) \left( 
\begin{array}{l}
F(1) \\ 
F(2) \\ 
F(3) \\ 
F(4)
\end{array}
\right)  \label{eq:nnnn}
\end{equation}
The combination of Eqs.(\ref{eq:gggg}), (\ref{eq:ffff}), and (\ref{eq:nnnn})
gives the expression for the second-order correlation function.

We first consider the coherent properties in the `number state' $|n\rangle
(n\neq 0).$ From Eq.(\ref{eq:gggg}), it is easy to obtain the second-order
correlation function

\begin{equation}
g^{(2)}=\frac{(n-1)(2j-n+2)}{n(2j-n+1)}.
\end{equation}
Then the `number state' $|j+1\rangle $ is always coherent as the
corresponding $g^{(2)}=1$, the states $|n\rangle (n>j+1)$ exhibit bunching,
and the states $|n\rangle (n<j+1)$ antibunching for integer $j$. 
Particularly $g^{(2)}=0$
for state $|1\rangle ,$ irrespective of $j.$ Next we numerically calculate $%
g^{(2)}$ for the SSCS.

Fig.1 gives the second-order correlation function of the SSCS for different
values of $\theta .$ We see that there exists a cross point for both the
integer and half-integer $j$ when $|\eta |=1.$ The coherence property at
this point is independent on the relative phase $\theta $. This is because
the two states $|\pm \eta \rangle $ are orthogonal with each other when $%
|\eta |=1.$ The orthogonality makes the correlation function be independent
of $\theta .$ For integer $j=3,$ we observe that $g^{(2)}(0)\geq g^{(2)}(\pi
/2)\geq g^{(2)}(\pi ).$ Here $g^{(2)}(\theta )\equiv g^{(2)}.$ However for
half-integer $j=5/2,$ $g^{(2)}(0)\geq g^{(2)}(\pi /2)\geq g^{(2)}(\pi )$
when $|\eta |<1$ and oppositely $g^{(2)}(0)\leq g^{(2)}(\pi /2)\leq
g^{(2)}(\pi )$ when $|\eta |>1.$ Obviously, the state $|\eta ,\pi \rangle $
exhibits high degree of antibunching. By comparing to the case of the SCS
(the corresponding correlation function is given by $g^{(2)}(\pi /2)$) the
antibuching effect is enhanced. In the limit of $|\eta |\rightarrow 0,$ the
state $|\eta ,\pi \rangle $ reduces to the number state $|1\rangle $ and the
correlation function is zero, which corresponds highest degree of
antibunching. In another limit $|\eta |\rightarrow \infty $ for integer $j,$
the limited states of the states $|\eta ,0\rangle $, $|\eta ,\pi /2\rangle $
and $|\eta ,\pi \rangle $ are $|2j\rangle ,|2j\rangle ,$ and $|2j-1\rangle ,$
respectively. Therefore the corresponding correlation functions are $1-j/2,$ 
$1-j/2,$ and $3(j-1)/(2j-1).$ For half-integer $j,$ the limiting states are $%
|2j-1\rangle ,|2j\rangle ,$ and $|2j\rangle $ and the limiting values of the
corresponding correlation functions are $3(j-1)/(2j-1),$ $1-j/2,$ and $1-j/2.
$ These limiting properties can also be seen from the figure.

From Eq.(\ref{eq:gggg}) we see that we can measure the second--order correlation
function by measuring the quantities $J_z^k(k=1...4)$. The correlation function is
dependent on the absolute value of $\eta$, i.e, it is phase--insensitive. However
the spin squeezing discussed below is a phase--sensitive quantity.

\begin{figure}
\epsfig{width=6cm,file=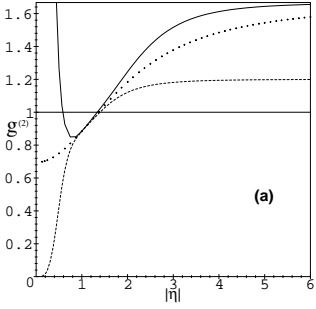}
\epsfig{width=6cm,file=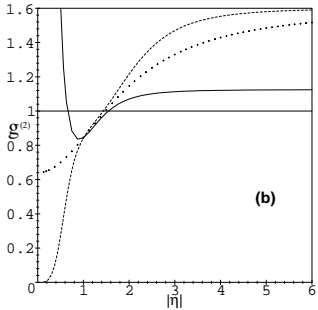}
\caption{The second-order correlation function as a function of $|\eta|$ for
different values of $\theta$. (a)integer $j=3$; (b)half-integer $j=5/2$. The
solid, dotted and dashed lines correspond to $\theta=0$, $\theta=\pi/2$ and $%
\theta=\pi$, respectively.}
\end{figure}

\section{Spin Squeezing}

In 1993 Kitagawa and Ueda showed that the spin squeezed state can be
produced by nonlinear Hamiltonians\cite{Kitagawa}. Later people develop
other ways to produce spin squeezing, such as by interaction of atoms with
squeezed light\cite{Wineland94,Kuzmich1,Lukin,Vernac}, quantum nondemolition
measurement of atomic spin states\cite{Kuzmichqnd}, and atomic collisional
interactions\cite{Soerensen1}.

In order to study spin squeezing we need to know some expectation values
such as $\langle J_{-}\rangle $ and $\langle J_{-}^2\rangle $ {\it et al.}
From the ladder operator formalism of the cat states(\ref{eq:laddercs2}),
the expectation value $\langle J_{-}\rangle $ is formally expressed as

\begin{equation}
\langle J_{-}\rangle =\eta \sqrt{\frac{1-\cos \theta \xi ^{2j}}{1+\cos
\theta \xi ^{2j}}}\langle \eta ,\theta |(2j-{\cal N})|\eta ,\theta +\pi
\rangle .  \label{eq:jjj}
\end{equation}
Then we need to know $\langle \eta ,\theta |\eta ,\theta +\pi \rangle $ and $%
\langle \eta ,\theta |{\cal N}|\eta ,\theta +\pi \rangle $ to determine $%
\langle J_{-}\rangle $ . We define a quantity $\tilde{G}(\lambda )=\langle
\eta ,\theta |\lambda ^{{\cal N}}|\eta ,\theta +\pi \rangle ,$ and obviously 
$\tilde{G}(1)=\langle \eta ,\theta |\eta ,\theta +\pi \rangle $ and $d\tilde{%
G}(\lambda )/d\lambda |_{\lambda =1}=\langle \eta ,\theta |{\cal N}|\eta
,\theta +\pi \rangle .$ From Eq. (\ref{eq:state}) the quantity $\tilde{G}%
(\lambda )$ is obtained as

\begin{equation}
\tilde{G}(\lambda )=\frac{-i\sin \theta (1-\lambda |\eta |^2)^{2j}}{\sqrt{%
1-\cos ^2\theta \xi ^{4j}}(1+|\eta |^2)^{2j}},
\end{equation}
which directly leads to

\begin{eqnarray}
\langle \eta ,\theta |\eta ,\theta +\pi \rangle &=&\frac{-i\sin \theta \xi
^{2j}}{\sqrt{1-\cos ^2\theta \xi ^{4j}}},  \nonumber \\
\langle \eta ,\theta |{\cal N}|\eta ,\theta +\pi \rangle &=&\frac{i2j\sin
\theta |\eta |^2\xi ^{2j}}{\sqrt{1-\cos ^2\theta \xi ^{4j}}(1-|\eta |^2)}.
\end{eqnarray}
Substituting the above equation into Eq.(\ref{eq:jjj}) we obtain

\begin{equation}
\langle J_{-}\rangle =\frac{-i2\eta j\sin \theta \xi ^{2j}}{\left( 1+\cos
\theta \xi ^{2j}\right) (1-|\eta |^2)}.  \label{eq:jminus}
\end{equation}

From Eqs.(\ref{eq:laddercs3}) and (\ref{eq:nnnn}), the expectation value of $%
J_{-}^2$ can be written in terms of the factorial moments as

\begin{equation}
\langle J_{-}^2\rangle =\eta ^2\left\{ F(2)-2(2j-1)\left[ F(1)-j\right]
\right\} .  \label{eq:jminus2}
\end{equation}

Having know the expectation values $\langle {\cal N}\rangle $ , $\langle 
{\cal N}^2\rangle ,$ $\langle J_{-}\rangle $ and $\langle J_{-}^2\rangle ,$
we can know the expectation values $J_\alpha $ and $J_\alpha
^2(\alpha=x,y,z) $ through the relations 
\begin{eqnarray}
J_x &=&\frac{J_{+}+J_{-}}2,\text{ }J_y=\frac{J_{+}-J_{-}}{2i},\text{ }J_z=%
{\cal N}-j,  \nonumber \\
J_x^2 &=&{\frac 14}[2j(2{\cal N}+1)-2{\cal N}^2+J_{+}^2+J_{-}^2],  \nonumber
\\
J_y^2 &=&{\frac 14}[2j(2{\cal N}+1)-2{\cal N}^2-J_{+}^2-J_{-}^2],  \nonumber
\\
J_z^2 &=&{\cal N}^2-2j{\cal N}+j^2.  \label{eq:jjjjj}
\end{eqnarray}
Then the squeezing parameter $\xi _x^2,$ $\xi _y^2,$ and $\xi _z^2$ are
obtained.

We study the spin squeezing in the even SCS $|\eta ,0\rangle $ and odd SCS $%
|\eta ,\pi \rangle $ and assume real parameter $\eta .$ From Eq.(\ref
{eq:jminus}) we know that the expectation value $\langle J_{-}\rangle =0$
for $\theta =0,\pi ,$ and then the expectation values $\langle J_x\rangle
=\langle J_y\rangle =0.$ Therefore the mean spin is along the $z$ direction.
The squeezing parameters , $\xi _x^2$ and $\xi _y^2$, now simplify to

\begin{equation}
\xi _x^2=\frac{2j\langle J_x^2\rangle }{\langle J_z\rangle ^2},\text{ }\xi
_y^2=\frac{2j\langle J_y^2\rangle }{\langle J_z\rangle ^2}.  \label{eq:xixi}
\end{equation}

First we examine the spin squeezing in the even SCS. Before performing
numerical calculations, we consider the case $j=1,$ which is the smallest $j$
that can make spin squeezing. From Eqs.(\ref{eq:ffff}-\ref{eq:nnnn}) and (%
\ref{eq:jminus2}-\ref{eq:xixi}), we obtain 
\begin{equation}
\xi _x^2=\frac{1+|\eta |^4}{(1-|\eta |^2)^2},\text{ }\xi _y^2=\frac{1+|\eta
|^4}{(1+|\eta |^2)^2}.
\end{equation}
The above equation shows that the even SCS always exhibits squeezing in the $%
y$ direction and no squeezing in the $x$ direction except two extreme values
of $|\eta |,$ $|\eta |=0,\infty .$ The maximum squeezing occurs at $|\eta |=1
$ and the corresponding squeezing parameter $\xi _y^2=1/2.$ These results
can also be seen from Fig.2.

Fig.2 gives a plot of inverse squeezing parameters $\xi _x^2$ and $\xi _y^2$
for different values of $j.$ First we see that the even SCS is not squeezed
in the $x$ direction, while in the $y$ direction, the state is squeezed.
Different from the case $j=1,$ the even SCS with $j>1$ is not always
squeezed. In the initial small range of $|\eta |,$ the state is squeezed and
becomes not squeezed when $|\eta |$ increases a little. The worst squeezing
occurs at $|\eta |=1$ for $j>1,$ oppositely the best squeezing occurs at the
same point for $j=1.$ When $|\eta |$ is large enough, the state with
half-integer $j$ is not squeezed, while the state with integer $j>1$ becomes
squeezed again. In the limit of $|\eta |\rightarrow \infty ,$ the squeezing
parameter $\xi _y^2$ is equal to 1 for integer $j$ and larger than 1 for
half-integer $j. $ 
\begin{figure}
\epsfig{width=8cm,file=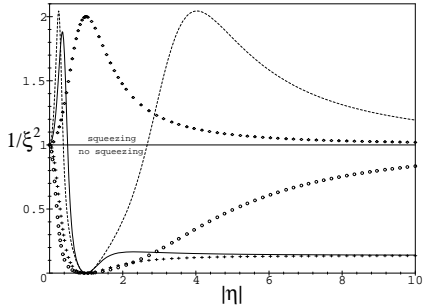}
\caption{ Inverse squeezing parameters $\xi_x^2$ and $\xi_y^2$ as a function
of $|\eta|$ for different $j$ in even SCS. The squeezing parameters and
parameter $j$ are choose as: $\xi_y^2$, $j=5/2$ (solid line), $j=5$(dashed
line), $j=1$(diamond points); $\xi_x^2$, $j=5/2$(cross points), $j=5$(circle
points).}
\end{figure}

Now we examine the squeezing properties in the odd SCS. The numerical
results are given by Fig.3. From the figure we observe that there is no
squeezing in the $x$ direction and even no squeezing in the odd SCS with
integer $j\,$along the $y$ direction$.$ The spin squeezing only exist in the
odd SCS with half-integer $j.\,$The state becomes squeezing after the
parameter $|\eta |$ across a critical point $|\eta _c|.$ After the critical
point the state is squeezed except the limit case $|\eta |\rightarrow \infty
.$ As half-integer $j$ increases we see that the value of $|\eta _c|$ and
the maximum value of $\xi _y^2$ increases, and the squeezing range decreases.

\begin{figure}
\epsfig{width=8cm,file=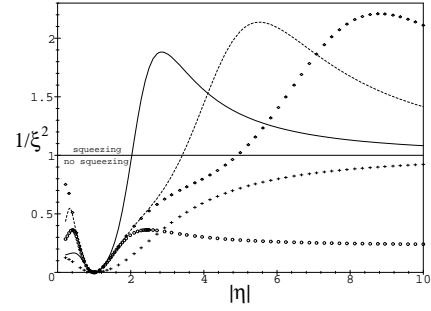}
\caption{ Inverse squeezing parameters $\xi_x^2$ and $\xi_y^2$ as a function
of $|\eta|$ for different $j$ in odd SCS. The squeezing parameters and
parameter $j$ are choose as: $\xi_y^2$, $j=5/2$ (solid line), $j=19/2$%
(dashed line), $j=49/2$(diamond points), $j=5$(circle points); $\xi_x^2$, $%
j=5/2$(cross points).}
\end{figure}

\section{Conclusion}

We have studied the coherence properties and spin squeezing in the SSCS. The
results show that the antibunching effect can be enhanced and spin squeezing
can be generated in the SSCS due to the superpositions of two SCSs. Both the
coherence properties and spin squeezing depend sensitively on the parity of $%
2j.$ Particularly the parity determines if the spin squeezing exists or not
in the odd SCS. 

In this paper we only consider the superpositions of two SCSs $|\pm\eta\rangle$.
It will be interesting to consider the spin squeezing in the superpositions of
more than two SCSs. As the spin squeezing  
is closely related the entanglement in multi-qubit systems, it will be helpful to 
understand entanglement by further investigation of spin squeezing.

\acknowledgments
The author thanks for many helpful discussions with Klaus M\o lmer and
Anders S\o rensen. This work is supported by the Information Society
Technologies Programme IST-1999-11053, EQUIP, action line 6-2-1 and the 
European Project Q-ACTA.


\begin{references}
\bibitem{EM}  For a review, see V. Bu\v {z}ek and P. L. Knight, in {\it %
Progress in Optics XXXIV}, edited by E. Wolf (Elsevier, Amsterdam, 1995).

\bibitem{Ion}  C. Monroe, D. M. Meckhof, B. E. King, and D. J. Wineland,
Science {\bf 272}, 1131 (1996).

\bibitem{RI}  V. V. Dodonov, S. Y. Kalmykov and V. I. Man'ko, V. I. Phys.
Lett. A{\bf 199}, 123 (1995).

\bibitem{SCS}  J. M. Radcliffe, J. Phys. A: Gen. Phys. {\bf 4}, 313 (1971);
F. T. Arecchi, E. Courtens, R. Gilmore, and H. Thomas, \pra {\bf 6}, 2211
(1972); R. Gilmore, C. M. Bowden and L. M. Narducci, \pra {\bf 12}, 1019
(1975); L. M. Narducci, C. M. Bowden, V. Bluemel, G. P. Garrazana, and R. A.
Tuft, \pra
{\bf 11}, 973 (1975); G. S. Agarwal, \pra {\bf 24}, 2889(1981).

\bibitem{Agar97}  G. S. Agarwal, R. R. Puri, and R. P. Singh, Phys. Rev. A 
{\bf 56}, 2246 (1997).

\bibitem{Gerry97}  C. C. Gerry and R. Grobe, Phys. Rev. A {\bf 56}, 2390
(1997).

\bibitem{Recamier}  J. Recamier, O. Castanos, R. J\'{a}uregui, and A. Frank,
Phys. Rev. A {\bf 61}, 063808 (2000).

\bibitem{Kitagawa}  M. Kitagawa and M. Ueda, \pra {\bf 47}, 5138 (1993).

\bibitem{Wineland94}  D. J. Wineland {\it et al}., \pra {\bf 46}, 11 (1992); %
\pra {\bf 46}, R6797 (1992); \pra {\bf 50}, 67 (1994).

\bibitem{Bg}  D. A. Trifonov, J. Math. Phys. {\bf 34}, 100 (1993); {\bf 35},
2297 (1994); Phys. Lett. A {\bf 187}, 284 (1994).

\bibitem{Soerensen3}  A. S\o rensen, L.-M. Duan, J. I. Cirac and P. Zoller,
Nature {\bf 409}, 63 (2001).

\bibitem{Wangspin}  X Wang,  J. Opt. B: Quantum and Semiclassical Optics 
{\bf 3},  93 (2001).

\bibitem{BS}  D. Stoler, B. E. A. Saleh, and M. C. Teich, Opt. Acta {\bf 32}%
, 345 (1985).

\bibitem{G2}  K. W\`{o}dkiewicz, Opt. Commun. {\bf 51}, 198 (1984); \prb 
{\bf 32}, 4750 (1985).

\bibitem{Kuzmich1}  A. Kuzmich, K. M\o lmer and E. S. Polzik, Phys. Rev.
Lett. {\bf 79}, 4782 (1997); A. Kozhekin, K. M\o lmer, and E. S. Polzik,
Phys. Rev. A {\bf 62}, 033809 (2000).

\bibitem{Lukin}  M. D. Lukin, S. F. Yelin, and M. Fleischhauer, Phys. Rev.
Lett.{\bf \ 84}, 4232 (2000).

\bibitem{Vernac}  L. Vernac, M. Pinard, and E. Giacobino, Phys. Rev. A {\bf %
62}, 063812 (2000).

\bibitem{Kuzmichqnd}  A. Kuzmich, N. P. Bigelow, and L. Mandel, Europhys.
Lett. {\bf 43}, 481 (1998); A. Kuzmich, L. Mandel, and N. P. Bigelow, Phys.
Rev. Lett. {\bf 85}, 1594 (2000).

\bibitem{Soerensen1}  A. S\o rensen and K. M\o lmer, Phys. Rev. Lett. {\bf 83%
}, 2274 (1999); X. Wang, A. S\o rensen and K. M\o lmer, Phys. Rev. A, to appear.
\end{references}
\end{document}